
\input phyzzx.tex
\pubnum{ROM2F-92-52}

\def\inte{ {1\over\pi}\int d^2\sigma\ }
\def\de{\partial}
\def\kin#1{\de_+ #1 \de_- #1}
\def\primo{^\prime}
\def\intx{{1\over 2\pi}\int d^2 x\ }
\def\kinx#1{\de_\mu #1\de^\mu #1}
\def\curv#1{i q_#1 R #1}
\def\unm{{1\over 2}}
\def\dmu#1{\mu{d #1\over d\mu}}
\def\bettyr#1{\beta_#1 := \dmu{#1^2_R}}
\def\boopr{\unm\gamma_{+R}\gamma_{-R}}

\def\lp{{\lambda^{(+)}}}
\def\lm{{\lambda^{(-)}}}
\def\lpm{{\lambda^{(\pm)}}}

\def\lsum{{\lp^2_R+\lm^2_R\over 2}}
\def\lmbb{\widehat{\lm}}
\def\lpbb{\widehat{\lp}}
\def\duc{2+(1-k)\kappa}
\def\radgR{\sqrt{-\hat g}\hat R}
\def\xop{x_0^+}
\def\mbh{m_{\rm bh}}

\titlepage
\title{A CONFORMAL AFFINE TODA MODEL OF 2D BLACK HOLES:\break
A QUANTUM STUDY OF THE EVAPORATION END-POINT }
\author{F.~Belgiorno, A.S.~Cattaneo\foot{Also Sezione
I.N.F.N. dell'Universit\`a di Milano, 20133 Milano, Italy}}
\address{Dipartimento di Fisica, Universit\`a di Milano\break
20133 Milano, Italy}
\author{F.~Fucito}
\address{Dipartimento di Fisica, Universit\`a di Roma II ``Tor Vergata"
and INFN sez. di Roma II\break
00173 Roma, Italy}
\andauthor{ M.~Martellini\foot{Permanent address: Dipartimento
di Fisica, Universit\`a di
Milano, 20133 Milano, Italy and Sezione I.N.F.N.
dell'Universit\`a di Pavia, 27100 Pavia, Italy}}
\address{Dipartimento di Fisica, Universit\`a di Roma I "La Sapienza",
00185 Roma, Italy}
\abstract
In this paper we reformulate the dilaton-gravity theory of Callan
\etal\ as
a new effective conformal field theory which turns out to be a generalization
of the so-called $SL_2$-conformal affine Toda (CAT) theory studied some
times ago by Babelon and Bonora. We quantize this model, thus keeping in
account the dilaton-gravity quantum effects.
We then implement a
Renormalization Group analysis to study the black hole thermodynamics
and the final state of the Hawking evaporation.
\endpage
\pagenumber=1
\chapter{Introduction}
\REF\CGHS{C.G.~Callan, S.B.~Giddings, J.A.~Harvey, and A.~Strominger
\journal Phys. Rev. &D45(92)R1005.}
\REF\BDD{T.~Banks, A.~Dabholkar, M.R.~Douglas, and M.~O'Loughlin
\journal Phys. Rev. &D45(92)3607.}
\REF\RST{J.G.~Russo, L.~Susskind, and L.~Thorlacius,
{\it Black Hole Evaporation in $1+1$ Dimensions},
Stanford preprint SU-ITP-92-4, January 1992;\nextline
L.~Susskind and L.~Thorlacius,
{\it Hawking Radiation and Back-Reaction},
Stanford preprint SU-ITP-92-12, hepth@xxx/9203054, March 1992.}
\REF\BGHS{B.~Birnir, S.B.~Giddings, J.A.~Harvey and A.~Strominger,
{\it Quantum Black Holes}, Santa Barbara preprint UCSB-TH-92-08,
hepth@xxx/9203042, March 1992.}
\REF\Str{A.~Strominger, {\it Faddeev--Popov Ghosts and $1+1$ Dimensional
Black Hole Evaporation},
UC Santa Barbara preprint UCSB-TH-92-18, hept@xxx/9205028, May 1992.}
\REF\Haw{S.W.~Hawking, {\it Evaporation of Two Dimensional Black Holes},
CALTECH preprint CALT-68-1774, March 1992.}
In a series of recent papers\refmark\CGHS$^{^-}$\refmark\Haw
a dilaton-gravity $2D$-theory\foot{The dilaton-gravity, coupled to $N$
matter fields is described by the following action:
$$
S=\intx\sqrt{-{\bf g}}\left[{
e^{-2\phi}(R+4(\nabla\phi)^2+4\Lambda^2)-\unm\sum_{i=1}^N(\nabla f_i)^2}
\right],
$$
where $\phi$ is the dilaton, $R$ the scalar curvature, $f$ the matter
fields and $\Lambda$ the cosmological constant, which is
supposed to be non-zero in order that the action have nontrivial solutions.}
which has
black hole solutions, known as the Callan--Giddings--Harvey--Strominger
(CGHS) model, was formulated in order to clarify the problem of the
black hole evaporation due to Hawking radiation.
Later Bilal and
Callan\Ref\BC{A.~Bilal and C.~Callan, {\it Liouville Models of Black Hole
Evaporation}, Princeton preprint PUPT-1320, hept@xxx/9205089, May 1992.}
and also de~Alwis\Ref\deA{S.P.~de~Alwis, {\it Black Hole Physics from
Liouville Theory}, Colorado preprint COLO-HEP-284, hepth@xxx/9206020,
June 1992.}
have reformulated the CGHS-model as a (but not exactly!)
Liouville field theory.
\par
However the resulting theory, which we shall call the Liouville-like
black hole (LBH) theory, is still ill-defined when the string coupling
constant $e^{2\phi}$ ($\phi$ is the dilaton) is of the order of a certain
critical value ${1\over\kappa}$, with $\kappa$ the coefficient of the
Polyakov kinetic term in the LBH-model. In this limit a singularity must
occur.
Furthermore the CGHS and LBH models are consistent if one could neglect
graviton-dilaton quantum effects.
\par
In this paper we shall show
that when $e^{2\phi}\sim{1\over\kappa}$ the LBH-model can be reformulated
as an exact theory generalizing the $SL_2$-conformal affine Toda theory,
previously studied by Babelon and
Bonora\Ref\BB{O.~Babelon and L.~Bonora \journal Phys. Lett. &B244(90)220.}
in the
framework of integrable models. In the following we shall refer to it as the
conformal affine Toda black hole (CATBH) model.
The CATBH-model allows a standard perturbative quantization
and in our picture such a quantization is just a device to unveil the
quantum effect of the dilaton $\phi$ and graviton $\rho$ because the
CAT-fields are suitable functions of ($\phi,\rho$).
This reformulation of the black hole evaporation problem has several
advantages:
\item{i)} It makes an investigation of the black hole physics around the
CGHS singularity possible, using the conformal field theory.
\item{ii)} We can study the final state of the black hole evaporation and
in particular the ``effective" black hole thermodynamics by applying
Renormalization Group (RG) techniques to the above conformal affine Toda
model.
By a physical point of view
the back-reaction should modify the Hawking radiation emission
and cause it to stop when the black hole has radiated away its initial ADM
mass. In our context the Hawking temperature is proportional to a certain
coupling constant, $\sqrt{\gamma_-}$, of the exponential interaction terms
of our CATBH-model. The key point here is to regard $\gamma_-$
as a RG-running
coupling constant in terms of the energy-mass scale $\mu$, which roughly
speaking measures the ADM mass. In particular if we formulate
an ansatz for the black hole solution interacting with $N$ infalling waves,
one may study the thermodynamics of the end point of the black hole evaporation
by extrapolating over RG-effective couplings to the (classical) energy scale
$t^\prime$ where the initial ADM-mass goes to zero. This picture gives an
approximate self consistent scheme to take into account the back reaction
problem in the mechanism of the Hawking radiation: the back reaction lies
in a dependence of the Hawking temperature over an energy scale $t$ and the
end point of the black hole evaporation is introduced by the limit
$t\to t^\prime$ where $M(t)\to M(t^\prime)\sim 0$. Here $M(t)$ is
the classical ADM-mass parametrized by $t$ and $M(t^\prime)\sim 0$
describes the end point state in which all the initial ADM-mass has been
radiated away.
\item{iii)} Since the CATBH-model is quantum integrable, by using the
bootstrap approach one may get the so-called quantum $R$-matrix and from
it one has, by standard techniques, the factorizable $S$-matrix. We shall
return on this argument in a forthcoming work\rlap.\Ref\BCMF{F.~Belgiorno,
A.S.~Cattaneo, M.~Martellini and F.~Fucito, {\it A Conformal Affine Toda
Model of Two Dimensional Black Holes: the End Point State and the
S-Matrix}, work in progress.}
\chapter{Conformal Affine Toda Black Hole Model}
Bilal  and Callan\refmark\BC\
have recently shown how to cast the CGHS black hole model
in a non standard Liouville-like form. Their key idea is to mimic here
the Distler--Kawai\Ref\DK{J.~Distler and H.~Kawai \journal Nucl. Phys.
&B321(89)509.}
approach to $2D$-quantum gravity so that one must take
into account in the CGHS model a dilaton dependent renormalization of the
cosmological constant as well as a Polyakov effective term induced by the
``total" conformal anomaly. The effect of the Polyakov term
in the ``flat" conformal gauge $\widehat{g_{\mu\nu}}
=-\unm e^{2\rho}\eta_{\mu\nu}$ will be to replace the coefficient ${N\over 12}$
of CGHS by a coefficient $\kappa$, where $\kappa$ will be fixed by
making the matter central charge (due to the $N$ free
scalar fields) cancel against the $c=-26$ diffeomorphism-ghosts contribution.
More specifically they were able to simplify the graviton-dilaton
action through a field redefinition of the form
$$
\matrix{
\omega = {e^{-\phi}\over \sqrt{\kappa}},
&\hfill
& \chi = {1\over 2} (\rho + \omega^2)},
\eqn\redoc
$$
where $\phi$ and $\rho$ are respectively the dilaton and the Liouville mode,
and $\kappa$ is a parameter of order $N$ to be adjusted by requirement of
conformal invariance.
According to this way of thinking $\kappa$ must be a free parameter.
However for $\kappa$
positive a singularity appears in the regime $\omega^2\to 1$,
which is also a strong coupling limit for small $\kappa$.
\par
Our purpose, as stated in the introduction, is to define an improvement of
the Bilal--Callan theory in such a way that a well defined exact conformal
field theory exists also in the ``phase" $\kappa$ positive and $\omega^2\sim
1$.
The surprise will be that such a theory must be a sort of ``massive
deformation", which must be conformal at the same time,
of the true Liouville theory,
as we shall see in the following.
\par
The kinetic part of the Bilal--Callan action is given by
$$
S_{\rm kin} = \inte\left[ {-4\kappa\kin{\chi} + 4\kappa (\omega^2 -1)
\kin{\omega}+{1\over 2} \sum_{i=1}^N \kin{f_i} }\right] ,
\eqn\bc
$$
and the energy-momentum tensor has the Feigin--Fuchs form
$$
T_{\pm\pm}=-4\kappa\de_\pm\chi\de_\pm\chi+2\kappa\de^2_\pm\chi
+4\kappa(\omega^2-1)\de_\pm\omega
\de_\pm\omega+{1\over 2}\sum_{i=1}^N\de_\pm f_i\de_\pm f_i.
\eqn\bcst
$$
When $\omega^2 >1$ the action and the stress tensor can be simplified
by setting
$$
\Omega = {1\over 2}\omega\sqrt{\omega^2-1}-
{1\over 2}\log{(\omega +\sqrt{\omega^2-1})} ,
\eqn\defom
$$
leading to the canonical kinetic term $4\kappa\kin{\Omega}$.
When $\omega^2 <1$ we must use the alternative definition
$$
\Omega\primo = {1\over 2}\omega\sqrt{1-\omega^2}-{1\over 2} \arccos{\omega} ,
\eqn\defompr
$$
obtaining the ``ghost-like" kinetic term $-4\kappa\kin{\Omega\primo}$.
We thus have two different theories describing different regions in the
space-time (remember that $\omega$ is a field), in the first one we have
a real positive valued field $\Omega$,
in the latter a ghost field $\Omega\primo$
with range in $[-{\pi\over 4},0]$. The first theory is the one we are
interested
in when considering the classical limit $\omega\to +\infty$
and has been studied in Ref. \BC.
If we do not restrict the range of $\omega$ from $1$ to
$\infty$, which is not a well defined  operation from a field theory
point of view, the
quantization of the Bilal--Callan sector (the one in which $\omega\in [1,
\infty]$) should provide an IR effective theory of the complete theory
in which $\omega$ has its natural range from $0$ to $\infty$.
\par
Particularly interesting is the region where $\omega^2(\sigma)-1$
is changing sign. Let us suppose that $\omega^2(\sigma_1)
> 1$ and $\omega^2(\sigma_2) < 1$, where $\sigma_1$ and
$\sigma_2$ are two very close points. Then our idea is to define an
improved Bilal--Callan kinetic action term in which both fields
$\Omega(\sigma_1)$ and $\Omega\primo(\sigma_2)$ appear. Namely we assume
the following contribution to such an improved action:
$$
4\kappa(\kin{\Omega(\sigma_1)}-\kin{\Omega\primo(\sigma_2)})
=\de_+\xi(\sigma)\de_-\eta(\sigma) +
\de_-\xi(\sigma)\de_+\eta(\sigma),
$$
where the new fields\foot{It is to be noted that the new fields $\eta$
and $\xi$ are limited from
below, but since the region of interest is around 0 this constraint is
significative only for $\xi$, which has to be positive.}
$$
\eqalign {
\eta(\sigma) &=\sqrt{2\kappa}C (\Omega(\sigma_1) + \Omega\primo(\sigma_2)),\cr
\xi(\sigma) &= {\sqrt{2\kappa}\over C} (\Omega(\sigma_1) -
\Omega\primo(\sigma_2)),\cr }
\eqn\defex
$$
are defined in the point $\sigma = {\sigma_1 + \sigma_2\over 2}$ and $C$ is
an irrelevant arbitrary constant.
\par
Let us now consider a field $\omega(\sigma)$ taking
values everywhere near to 1
and rapidly fluctuating  up and down. We can
renormalize (\'a la Wilson) the above theory in which the undefined
Bilal--Callan kinetic term
$(\omega^2-1)\kin{\omega}$ has been replaced by the undefined kinetic term
$\de_+\xi\de_-\eta +\de_-\xi\de_+\eta$. From a naive dimensional argument
when $\omega^2\sim 1$ the Laplacian term $\kin\omega$ must be very large
in order to have a non trivial propagator for the $\omega$ field.
This means that our theory is a sort of ``UV effective theory" for the LBH
model.
\par
The next step is to identify the cosmological constant term which can be
added to our improved kinetic term. Our strategy here will be the same
one of Ref. \BC, but with a subtle difference: the cosmological
constant action must be identified with a potential of
conformal weight $(1,1)$ with respect to the improved stress tensor
$T(\sigma)={T(\sigma_1)+T(\sigma_2)\over 2}$ (where $T(\sigma_{1,2})$ is
defined in \bcst), which becomes:
$$
T_{\pm\pm}=-4\kappa\de_\pm\chi\de_\pm\chi+2\kappa\de^2_\pm\chi
+2\de_\pm\xi\de_\pm\eta+2C^2q\de_\pm^2\xi-2q\de_\pm^2\eta
+{1\over 2}\sum_{i=1}^N\de_\pm f_i\de_\pm f_i,
\eqn\ourst
$$
where we have introduced ``arbitrary" improvement terms also for $\xi$
and $\eta$ which, turning back to $\Omega$ in the LBH-model,
vanish as a consequence of \defex.
\par
If we choose to consider operators of exponential form, we guess that our
complete (\ie\ kinetic plus cosmological constant) improved action reads:
$$
S = \intx\left[{-\kappa\kinx\chi+\de_\mu\eta\de^\mu\xi
+\gamma_+ e^{A_+\chi}+\gamma_- e^{-A_-\chi + i\delta\eta}}\right] ,
\eqn\impract
$$
where $A_\pm$ are constrained by requiring conformal invariance and
$\delta$ is a free parameter since it contributes to the weight of the operator
only through the combination $q\delta$ (see \ourst) and $q$ can
always be conveniently adjusted.
\par
Even if this is not the most general action constructed with conformal
perturbations of weight (1,1) with respect to the improved
stress tensor \ourst\ (as other exponential combinations of the fields
$\chi$, $\eta$ and $\xi$ are possible), one should show
that \impract\ is actually sufficient\rlap.\refmark\BCMF
Very recently Giddings and Strominger\Ref\GS{S.B.~Giddings and
A.~Strominger, {\it Quantum Theory of Dilaton Gravity}, UC Santa Barbara
preprint, UCSB-TH-92-28, hepth@xxx/9207034, July 1992.}
have argued that there is an infinite number of quantum theories of
dilaton-gravity and the basic problem is to find physical criteria to narrow
the class of solutions. Our approach here is to specify the class of solutions
of cosmological constant action, which may give a known soluble (for us
this means integrable) conformal field theory with physical behaviours.
This choice exists and is the one giving \impract, since if
we perform the redefinition
$$
\varphi = i\sqrt{2\kappa}\chi,
\eqn\defvarp
$$
we obtain the well defined conformal invariant
Babelon--Bonora-like theory\refmark\BB
$$
S_{UV}=\intx\left[{{1\over 2}\kinx{\varphi}+\de_\mu\xi\de^\mu\eta
+\gamma_+ e^{i\lambda^{(+)}\varphi}+\gamma_- e^{-i\lambda^{(-)}\varphi
+i\delta\eta}}\right],\eqn\bblike
$$
where $\lambda^{(\pm )}=-{A_\pm\over\sqrt{2\kappa}}$.
At the classical level one may always set $\gamma_+=\gamma_-=\gamma$
and $\delta=\lm=\lp=\lambda$ ($\delta$ is a free parameter) in
\bblike; in the following we understand in this sense the classical
limit of~\bblike.
\par
Notice that this is the unique action in terms of three fields which is
at the same time classically integrable, \ie\ admitting a Lax pair,
and conformally invariant as shown by Babelon and Bonora.
\par
In the ``black hole physics limit" the interaction vertex
proportional to $\gamma_-$ in \bblike\ should give
the expected correct behaviour
of the cosmological part of the Bilal--Callan action and
the $\gamma_+$-vertex should become negligible in
the same limit. We shall see in the next section that, using a RG argument,
this behaviour can be actually obtained at a suitable scale of energy.
\chapter{Renormalization Group Analysis}
We want to consider here the renormalization flow of the classical
Babelon-Bonora action
$$
S_{BB}=\intx\left[{{1\over 2}\kinx{\varphi}+\de_\mu\eta\de^\mu\xi
-2\left({e^{2\varphi}+e^{2\eta-2\varphi}}\right) }\right].
\eqn\bb
$$
At the quantum level one must implement wave and vertex function
renormalizations so that in \bb\ one must introduce different
bare coupling constants in front of the fields as well as in front of
the vertex interaction terms. As a consequence one ends with the form
\bblike. However, according to the general spirit of the
renormalization procedure
all generally covariant dimension 2 counter terms are possible in \bblike\
and hence also Feigin--Fuchs terms, \ie\ the ones involving the $2D$-scalar
curvature. This ansatz is in agreement with the perturbative theory as
one could show following Distler and Kawai.
In our context the Feigin--Fuchs terms come naturally out if we look at
the improved stress tensor \ourst.
This leads us to consider the following
generalized form of the BB action in a curved space:
$$
\eqalign {
S &=\intx\sqrt{ g}\left[ {
g^{\mu\nu} \left({{1\over 2}\de_\mu\varphi\de_\nu\varphi +
\de_\mu\eta\de_\nu\xi}\right)
+\gamma_+e^{i\lp\varphi}+\gamma_-e^{i\delta\eta-i\lm\varphi}
}\right.\cr
&+\curv\varphi +\curv\eta +\curv\xi ] .\cr }
\eqn\gbb
$$
\par
Let us now start with the renormalization procedure of
\gbb\ in a perturbative framework, in the hypothesis that
$\lm^2\sim\lp^2\sim4$.
Notice that $\xi$ plays the role of an auxiliary field, a variation with
respect to which gives the on-shell equation of motion
$$
\nabla_\mu\nabla^\mu\eta=iq_\xi R,
\eqn\consh
$$
which in our perturbative scheme must be linearized around the
flat space, giving:
$$
\de_\mu\de^\mu\eta=0.
\eqn\onsh
$$
\REF\GLPZ{M.T.~Grisaru, A.~Lerda, S.~Penati and D.~Zanon
\journal Nucl. Phys. &B342(90)564.}
Following Ref.~\GLPZ, we define the renormalized quantities
at an arbitrary mass scale $\mu$ by:
$$
\matrix{
\eqalign{
\varphi &= Z^\unm_\varphi \varphi_R,\cr
\eta &= Z^\unm_\eta \eta_R,\cr
\xi &= Z^{-\unm}_\eta \xi_R,\cr}
&\eqalign{
\gamma_\pm &= \mu^2 Z_{\gamma_\pm} \gamma_{\pm R},\cr
\lpm^2 &= Z^{-1}_\varphi \lpm^2_R,\cr
\delta^2 &= Z^{-1}_\eta \delta^2_R,\cr}
&\eqalign{
 q^2_\varphi &= Z^{-1}_\varphi q^2_{\varphi R},\cr
 q^2_\eta &= Z^{-1}_\eta q^2_{\eta R},\cr
 q^2_\xi &= Z_\eta q^2_{\xi R}.\cr }
\cr}
\eqn\renquant
$$
The following quantities are conserved through renormalization:
$$
\matrix {
r={q_\varphi\over\lm},
& l={\lp\over\lm},
& k=q_\xi\delta,
& p=q_\xi q_\eta.\cr }
\eqn\constquant
$$
Following essentially the same procedures of Ref.~\GLPZ, with slightly
modifications due to the presence of extra fields and
by taking $\lp\not=\lm$, we
find that the theory can be renormalized at one loop
if we restrict ourself to the
on-shell renormalization scheme, \ie\ if we get rid of the terms in
$\de_\mu\de^\mu\eta$, produced by the renormalization, using
\onsh\rlap.\foot{The true on-shell theory
should rely on \consh, but at this perturbative order curvature terms
can be neglected (they are important only in the
renormalization of $\gamma_\pm$).} The curvature terms are taken into
account\Ref\Zam{A.B.~Zamolodchikov
\journal JEPT Lett. &43(86)730.}
considering the modifications to the trace of the stress tensor,
calculated on-shell. We finally get the following $\beta$ functions:
$$
\eqalign {
\beta_+:=\dmu{\gamma_{+R}}
&= 2\gamma_{+R}\left({{\lp^2_R\over 4}-1-q_{\varphi R}\lp_R}\right),\cr
\beta_-:=\dmu{\gamma_{-R}}
&= 2\gamma_{-R}\left({{\lm^2_R\over 4}-1+q_{\varphi R}\lm_R-q_{\xi R}
\delta_R}\right),\cr}
\eqn\betag
$$
$$
\eqalign{
\bettyr{\lpm} &= \boopr{\lp^2_R+\lm^2_R\over2}\lpm^2_R,\cr
\bettyr\delta &= \boopr\delta^4_R,\cr}
\eqn\betal
$$
$$
\eqalign{
\bettyr{{q_\varphi}} &= \boopr\lsum q^2_{\varphi R},\cr
\bettyr{{q_\eta}} &= \boopr\delta^2_R q^2_{\eta R},\cr
\bettyr{{q_\xi}} &= -\boopr\delta^2_R q^2_{\xi R}.\cr }
\eqn\betas
$$
Putting together the equations \betal, we find that also the following
quantity is conserved through renormalization\rlap:\foot{This result,
deriving from \betal, need not be valid out of this perturbative order,
while it is obviously true for the quantities in \constquant.}
$$
d={1+l^2\over 4}{1\over\delta^2}-{1\over\lambda^2},
\eqn\constquantd
$$
\par
Using the non-perturbative RG-invariants in \constquant\ and dropping out
the rather cumbersome $R$ indices, we can rewrite the relevant
$\beta$ functions as:
$$
\eqalign {
\beta_+ &=2\gamma_+\left({
{l^2-4lr\over 4}\lm^2-1}\right)\cr
\beta_- &=2\gamma_-\left({
{1+4r\over 4}\lm^2-1-k}\right)\cr
\beta_\lm &= {1+l^2\over 4}\gamma_+\gamma_-\lm^4,\cr}
\eqn\betasimple
$$
With the position
$$
\alpha= {1+l^2+4r(1-l)\over 2},
\eqn\defar
$$
we can solve \betasimple, for $\lm^2\sim4$, obtaining
$$
t-t_0={1\over\alpha b}\log\left|{
{\lambda^2_- -\lm^2\over\lm^2-\lambda^2_+}}\right|,
\eqn\sollambdad
$$
where $t=\log\mu$ and $b$ is an integration constant supposed to be
real in order to have RG fixed points, which are located at:
$$
\lambda^2_\pm={4+2k\over\alpha}\mp b.
\eqn\RGfixd
$$
Writing explicitly \sollambdad\ and solving \betasimple\ also for $\gamma_\pm$,
we eventually find\rlap:\foot{Even if the RG equations admit another
solution, \ie\ $\lm\not\in[\lambda_+,\lambda_-]$, we consider here
only the solution with
$\lambda_+^2<\lm^2<\lambda_-^2$, which is in agreement, as we shall
see, with our limiting case provided by the LBH-model.}
$$
\eqalign{
\lm^2 &={2k+4\over\alpha}-b\tanh{\alpha b(t-t_0)\over 2},\cr
\gamma_-^2&=Ae^{(8\tilde rk+16\tilde r-2k)(t-t_0)}\left[{
\cosh{\alpha b(t-t_0)\over 2}}\right]^{-2-8\tilde r},\cr
\gamma_+^2&=Be^{-(8\tilde rk+16\tilde r-2k)(t-t_0)}\left[{
\cosh{\alpha b(t-t_0)\over 2}}\right]^{-2+8\tilde r},\cr}
\eqn\solin
$$
where
$$
\tilde r= {1-l^2+4r(1+l)\over 8\alpha}.
\eqn\defr
$$
\par
We have described so far the most general situation in which all the
parameters are unconstrained. Indeed, following the reasoning of Sec.~2,
we have to request that at a certain scale $\hat t$, which shall be
proved to exist, both vertex operators have a conformal weight
$(1,1)$. This is achieved imposing the following
constraints\rlap:\foot{Notice that imposing these conditions is equivalent to
asking that, at the scale $\hat t$, both $\beta_\pm$ in \betag\
vanish.}
$$
\eqalign{
&{\lpbb^2\over 4}-r\lmbb\lpbb=1,\cr
&\left({ {1\over 4}+r }\right)\lmbb^2-k=1,\cr}
\eqn\constr
$$
where $\hat \lambda$ stands for the running coupling constant
at the scale $\hat t$.
Solving \constr\ with respect to the variables
$\widehat{\lpm}$, we obtain:
$$
\eqalign{
\lmbb &= \pm2\sqrt{1+k\over 1+4r},\cr
\lpbb &= 2(r\lmbb\pm\sqrt{r^2\lmbb^2+1}).\cr}
\eqn\BBsol
$$
We can now calculate the values of the constants $\alpha$ and $\tilde r$
in \defar\ and \defr\ as functions of the parameters $k$ and $r$:
$$
\eqalign{
\alpha &= {(1+4r)(2+k)\over 2(1+k)},\cr
\tilde r &= {k\over 4(2+k)}.\cr}
\eqn\arsol
$$
As a consequence of this, ${2k+4\over\alpha}$
is equal to $\lmbb^2$ for any value of $k$
and $r$. This means by \solin\ that
the scale $\hat t$ is just the renormalization scale $t_0$.
\par
It is also easily seen that the particular combination $8\tilde rk+16
\tilde r-2k$ vanishes identically for any value of $k$, so that the
exponential parts in the $\gamma_\pm$ scaling laws \solin\ are missing.
This implies that $\gamma_\pm$ depend on the scale $t-t_0$ only through
the cosh, which is an even function, and hence do not distinguish
between IR and UV scales. The requirement of having vertex
operators with the right conformal weight at the defining scale $t_0$
forces the theory to be dual (under the exchange
of the IR and UV scales).
\par
Considering what must happen at the scale $t_{BC}$, where the theory becomes
Bilal--Callan--like, and using \defvarp\ we find that
$$
\matrix{
q_\varphi(t_{BC})=\sqrt{\kappa\over 2}\hfill
&\hbox{and}
&\hfill\lm(t_{BC})=\sqrt{8\over\kappa}.\cr}
\eqn\qlBC
$$
By \constquant\ and \qlBC\ we get
$$
r={\kappa\over 4},
\eqn\rsol
$$
and hence, from \arsol we get:
$$
\alpha={(1+\kappa)(2+k)\over 2(1+k)}.
\eqn\asol
$$
Finally we want the $\eta$-independent vertex operator disappear
--- remember that $\eta$ and $\xi$ go to $\Omega$ in some ``classical"
limit --- and hence we must impose that
$$
\left|{{\gamma_-(t_{BC})\over\gamma_+(t_{BC})}}\right| >> 1.
\eqn\rap
$$
This implies that the following conditions must hold:
\item{i)}$\tilde r  < 0$, which
can also be read, by \arsol, as $-2<k<0$. This condition
can be further restrained to lead to $-1<k<0$,
as $k<-1$ would imply an imaginary $\lmbb$\rlap.\foot{Notice here
the necessity of having $\lp\not=\lm$,
since otherwise we would have $\tilde r=r$ by \defr\
and the condition would not be satisfied in any way.}
\item{ii)}$|\alpha b(t_{BC}-t_0)| >> 1$, which implies that $\lm(t_{BC})$ must
be substantially equal to the asymptotic value $\lambda_\pm$, given by
\RGfixd.
\par\noindent
We can solve the second constraint with respect to $b$, getting
$$
b=4{2+(1-k)\kappa\over\kappa(1+\kappa)},
\eqn\eqforb
$$
and now, also using \solin, \arsol\ and \rsol,
we can reexpress the running coupling constant $\lm(t)$ and
$\gamma_-(t)$ as:
$$
\eqalign{
\lm^2(t) &= {8\over2+\kappa}\left\{{
1+k-{\duc\over\kappa}\tanh\left[{
{\duc\over\kappa}{2+k\over1+k}(t-t_0)}\right]}\right\},\cr
\gamma_-(t) &= A'\left\{{
\cosh\left[{
{\duc\over\kappa}{2+k\over1+k}(t-t_0)}\right]}\right\}^{-2{1+k\over2+k}}.\cr}
\eqn\newsolin
$$
The duality implies that $\gamma_-(t)$ behaves in the same manner both at
an UV and at an IR scale; indeed we find, from \newsolin, the asymptotic
behaviour:
$$
\matrix{
\gamma_-(t){\buildrel |t-t_0|\to\infty \over\sim}
A'e^{-2s|t-t_0|},\hfil
&\hbox{ with }
&\hfil s={2+(1-k)\kappa \over\kappa}.\cr}
\eqn\defs
$$
\par
Let us eventually consider the remaining boundary conditions.
Since by \ourst\ $q_\xi=-q_\eta$, in order to have a correspondence with
Ref.\BC, we must set, using \defex, $\delta=4i/\sqrt{2\kappa}$, and
$p=k^2\kappa/8$. The vanishing of the total central charge
$c_{tot}$, which in the free theory can be calculated to be equal to
$N-23-24q_\varphi^2-48p$, at the scale $t_{BC}$ implies by the first
equation of \qlBC\ that
$\kappa=(N-23)/(12+6k^2)$. A comparison with the
asymptotic relation $\kappa\buildrel N\to\infty\over\sim N/12$, tells us
that $k\to0^-$ in the same limit, so that by \defs:
$$
s\buildrel N\to\infty\over\longrightarrow 1.
\eqn\lims
$$
\chapter{Black Hole Thermodynamics}
In this section we want to give a physical interpretation of the results
of our RG analysis.
We start from the observation made in Ref.~\Haw\
that, in the 2D-black hole described by the CGHS model, the Hawking temperature
$T_H$ is proportional to the cosmological constant, which in our context is
represented by $\sqrt{\gamma_-(t_{BC})}$. We can thus think of the
CATBH running coupling constant $\gamma_-(t)$ as of a scale dependent Hawking
temperature $T_H(t)$:
$$
T_H(t)\propto \sqrt{\gamma_-(t)}.
\eqn\eqforTH
$$
\par
A relation between the v.~e.~v. of the operator-valued
scalar curvature $\sqrt{-g} R$ and the other CATBH running coupling constant
$\lm$ can be obtained in the classical limit $e^\phi\to 0$.
Indeed, if we consider the
conformal gauge $\widehat{g_{\mu\nu}}=-\unm e^{2\lambda\rho}\eta_{\mu\nu}$,
we get in tree approximation:
$$
<\sqrt{-\hat g}\hat R> = 2\lambda\de_\mu\de^\mu<\rho>.
\eqn\classeqforgR
$$
\par
We may then state something about the end point state of the black hole
evaporation, for the explicit solution describing the black
hole formation by $N$-shock waves $f_i$, with $<\rho>$ computed
at the tree level.
In our contest, the CGHS ansatz for the classical solution $<\rho>\equiv
<\rho>_{\rm tree}$ looks like:
$$
\eqalign{
e^{-2\lm(t)<\rho(x^+)>} &=
1-2\lm(t)<\rho(x^+)>+O(\lm^2)\cr
&=-\kappa a(x^+-x_0^+)\theta(x^+-x_0^+)-\gamma_-(t)x^+x^-,\cr}
\eqn\sette
$$
where $\theta$ is the Heaviside function and
$x^\pm\equiv x^0\pm x^1$ and $a\equiv{\rm const}$.
Therefore, at $x^+=x^+_0$, where the $f$-waves are sitting,
we have by \classeqforgR, using the light-cone
coordinates $x^\pm$:
$$
\matrix{
<\sqrt{-\hat g(x^+)}\hat R(x^+)>=
\de_+\de_-[2\lm(t)<\rho(x^+)>]\sim\gamma_-(t),\hfill
&\hfill x^+\to x_0^+.\cr}
\eqn\otto
$$
Since here we have two scales $x^+_0$ and $\mu$,
it is reasonable to set (in $c=\hbar=1$) $x^+_0\equiv{1\over\mu}$, since
the natural scale which describes the black hole formation is $\xop$.
Therefore we arrive to the ``phenomenological" (in the sense it is based
on a suitable on-shell solution) relation:
$$
<[\radgR](t)>\propto\gamma_-(t).
\eqn\nove
$$
But now we know that the
arbitrary RG-energy scale $t$ is connected to the black hole evolution length
scale $\xop$ by the simple relation $t\equiv\log(1/\xop)$.
Since in the CGHS ansatz the black hole mass $\mbh$ grows linearly
with $\xop$, \ie\ $\mbh\propto M^2_{\rm Pl}\xop$ with $M_{\rm Pl}$ the
Planck mass, \defs\ reads now, using \eqforTH:
$$
T_H(\mbh)\buildrel\mbh\to0\over\sim T_0\left({\mbh\over m_0}\right)^s,
\eqn\mtozero
$$
and
$$
T_H(\mbh)\buildrel\mbh\to\infty\over\sim T_0\left({m_0\over\mbh}\right)^s,
\eqn\mtoinfty
$$
where $T_0$ and $m_0$ are arbitrary constants.
The vanishing of the Hawking temperature for small and large black hole
masses is a consequence of the duality between the UV and IR limit of the
quantum theory, which connects the small to the large mass scale.
For $N\to\infty$ we have
by \lims\ $s\to1$ and  thus we recover Hawking's semiclassical
result for the astrophysical black hole $\mbh\to\infty$. This is a very
nice feature that supports the self-consistency of our picture.
\par
The end point state of the black hole evaporation is characterized by the
limit $\mbh\to0$. But in this limit $T_H$ and by \nove\ also
$<\radgR>$ are vanishing. We understand this result
as a signal that at the end point the black hole \underbar{disappears
completely} from our $2D$-universe, leaving a zero temperature regular
remnant solution. This scenario has been suggested by
Hawking\Ref\Hawbis{S.W.~Hawking, \journal Phys. Rev. &D14(76)2460.}
and 't~Hooft\rlap,\Ref\tH{G.~t'Hooft \journal Nucl. Phys. &B335(90)138.}
but with a basic difference: for Hawking ('t~Hooft) the final
state is mixed (pure). Of course at the level of the above RG analysis we
cannot say anything on the final Hilbert space. However, we have an explicit
quantum field model for answering, in principle, to the above question.
Our point of view is to see whether the $S$-matrix associated with
the ``quantum"
black hole states and coming from the scattering operator corresponding to
our CAT model \gbb\ is unitary or not. Clearly a unitary $S$-matrix may be
in agreement only with 't~Hooft scenario. In a forthcoming paper,
Ref.~\BCMF, we shall consider this problem in full detail.
\ack
One of us (M.M.) wants to thank A.Ashtekar and L.Bonora for illuminating
discussions.
\endpage\refout
\bye